%% file: ms.tex
\documentclass{new_tlp}
\pdfoutput=1
\input{header}
\begin{document}
\selectlanguage{english}
\input{Deckblatt}
\input{Chap1_Introduction}
\input{Chap2_Background}
\input{Chap3_OptimalPUs}
\input{Chap4_OptimalTime}
\input{Chap5_OptimalFurthers}

\input{Chap6_Benchmarks}
\input{Chap7_RelatedWork}

\input{Chap8_Conclusion}

%list all bib entries (incl nonreferenced)
%\nocite{*}

%create list of literature
%\bibliography{bibtex/Literaturverzeichnis}{}
\bibliography{ms}{}
\bibliographystyle{sty/acmtrans}

\label{lastpage}
%\clearpage

%\input{Appendix}
\end{document}

%% file: header.tex
\usepackage{mathptmx}                   % Math and co from new_tlp
\usepackage[utf8]{inputenc}             % Encoding UTF8
\usepackage[english,ngerman]{babel}     %import languages
\usepackage{layout}                     %enable layout debugging with \layout
\usepackage{hyperref}                   %enable hyperrefs
\hypersetup{
  pdfinfo={
    Title={Optimal Scheduling for Exposed Datapath Architectures with Buffered Processing Units by ASP},
    Author={Marc Dahlem, Anoop Bhagyanath, and Klaus Schneider},
    Keywords={Answer Set Programming, Exposed Datapath Architectures, Optimal Scheduling, Code Generation}
  }
}
\usepackage{tabto}                      %enable tabto
\usepackage[bottom,hang]{footmisc}      %Fußnoten
\usepackage{tikz}                       %tikz
\usetikzlibrary{shadows}
\usepackage{float}
\usepackage{sty/eslistings}
\addto\extrasenglish{%

}

\usepackage{graphicx}
\usepackage{subcaption}
\usepackage{wrapfig}
\usepackage{url}
\usepackage{amssymb}
\usepackage{sty/mathpartir}
\usepackage{etoolbox}
\patchcmd{\quote}{\rightmargin}{\leftmargin 1em \rightmargin}{}{}

\newcommand\bcmdtab{\noindent\bgroup\tabcolsep=0pt%
  \begin{tabular}{@{}p{10pc}@{}p{20pc}@{}}}
\newcommand\ecmdtab{\end{tabular}\egroup}

\pagerange{\pageref{firstpage}--\pageref{lastpage}}
\doi{}

%% file: Deckblatt.tex
\title[Optimal SCAD-Scheduling by ASP]{Optimal Scheduling for Exposed Datapath Architectures with Buffered Processing Units by ASP}

%%%%%%%%%%%%%%%%%%%%%%%%%%%%%%%%%%%%%%%%%%%%%%%%%%%%%%%%%%%%%%%%%%%%%%%%%%%%%%%%
\author[Marc Dahlem, Anoop Bhagyanath, and Klaus Schneider]
    {Marc Dahlem\\
    Insiders Technologies GmbH, Kaiserslautern, Germany\\  https://insiders-technologies.de
    \and Anoop Bhagyanath, Klaus Schneider\\
    Department of Computer Science, University of Kaiserslautern, Germany\\https://es.cs.uni-kl.de
}
%%%%%%%%%%%%%%%%%%%%%%%%%%%%%%%%%%%%%%%%%%%%%%%%%%%%%%%%%%%%%%%%%%%%%%%%%%%%%%%%
\jdate{February 2018}
\pubyear{2018}

\label{firstpage}
\maketitle

\begin{abstract}
Conventional processor architectures are restricted in exploiting instruction level parallelism (ILP) due to the relatively low number of programmer-visible registers. Therefore, more recent processor architectures expose their datapaths so that the compiler (1) can schedule parallel instructions to different processing units and (2) can make effective use of local storage of the processing units. Among these architectures, the Synchronous Control Asynchronous Dataflow (SCAD) architecture is a new exposed datapath architecture whose processing units are equipped with first-in first-out (FIFO) buffers at their input and output ports.

In contrast to register-based machines, the optimal code generation for SCAD is still a matter of research. In particular, SAT and SMT solvers were used to generate optimal resource constrained and optimal time constrained schedules for SCAD, respectively. As Answer Set Programming (ASP) offers better flexibility in handling such scheduling problems, we focus in this paper on using an answer set solver for both resource and time constrained optimal SCAD code generation. As a major benefit of using ASP, we are able to generate \emph{all} optimal schedules for a given program which allows one to study their properties. Furthermore, the experimental results of this paper demonstrate that the answer set solver can compete with SAT solvers and outperforms SMT solvers.
\emph{This paper is under consideration for acceptance in TPLP.}
\end{abstract}

\begin{keywords}
Answer Set Programming, Exposed Datapath Architectures, Optimal Scheduling, Code Generation
\end{keywords}

%\tableofcontents

%% file: Chap1_Introduction.tex
%-------------------------------------------------------------------------------
\section{Introduction} \label{chap:introduction}
%-------------------------------------------------------------------------------
%-------------------------------------------------------------------------------
\subsection{Motivation}
%-------------------------------------------------------------------------------
\noindent Memory access has always been a critical issue for improving the execution time of programs for at least the following reasons: First, the relative improvement of memory access time has not been as good as the relative improvement of execution time of other instructions of processors. Second, the sequential access to the main memory limits the availability of data for concurrent executions by functional units in processors. The former is tackled to some extent by the development of expensive memory hierarchies starting with the registers of the processor itself. Storing intermediate values of a program in registers allows processors to access these values much faster compared to the next components in the memory hierarchy. However, the number of registers of a register file and number of its ports, i.e., the number of parallel reads/writes of registers, are usually quite low. Increasing these numbers is difficult since the number of registers is directly encoded in the instruction sets. Changing it requires corresponding changes in processors and compilers. Also, increasing the number of ports of register files and processing units (PUs) quickly leads to a bottleneck in wiring these on the chips \cite{ACGK07}. However, a small number of ports of register files quickly becomes a communication bottleneck in processors since all values are written to and read from registers. This is particularly true for recent processors that are employing many functional units. Therefore, while introducing registers was a good idea for sequential processors, it now limits the use of instruction level parallelism (ILP) for recent architectures.

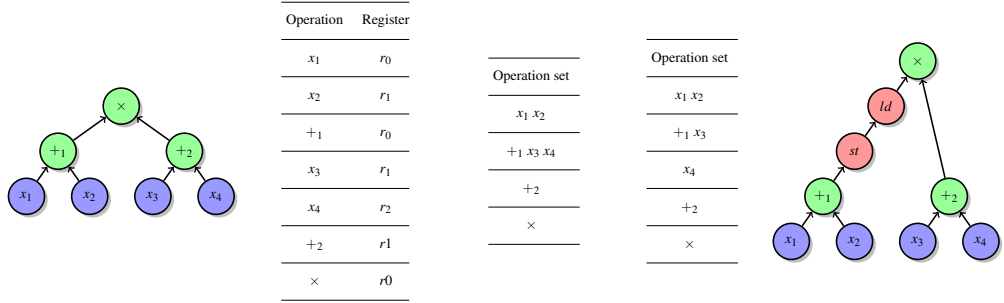
\begin{figure}
\begin{subfigure}{0.22\textwidth}
\centering
\scalebox{0.6}{%
\begin{tikzpicture}[xscale=0.7,yscale=0.5]
\pgfsetlinewidth{1pt}
\tikzstyle{vr}=[circle,minimum size=8mm,draw,fill=blue!40,drop shadow]
\tikzstyle{op}=[circle,minimum size=8mm,draw,fill=green!40,drop shadow]
\draw (0,2) node[vr] (x1) {$x_1$};
\draw (2,2) node[vr] (x2) {$x_2$};
\draw (4,2) node[vr] (x3) {$x_3$};
\draw (6,2) node[vr] (x4) {$x_4$};
\draw (1,4) node[op] (y1) {$+_1$};
\draw (5,4) node[op] (y2) {$+_2$};
\draw (3,6) node[op] (z1) {$\times$};
\draw[->](x1) to (y1);
\draw[->](x2) to (y1);
\draw[->](x3) to (y2);
\draw[->](x4) to (y2);
\draw[->](y1) to (z1);
\draw[->](y2) to (z1);
\end{tikzpicture}}
\end{subfigure}
\begin{subfigure}{0.21\textwidth}
\centering
\scalebox{0.6}{
\begin{tabular}{cc}
  \hline
  Operation & Register \\
  \hline
  $x_1$ & $r_0$ \\ 
  \hline
  $x_2$ & $r_1$ \\
  \hline
  $+_1$ & $r_0$ \\ 
  \hline
  $x_3$ & $r_1$ \\
  \hline
  $x_4$ & $r_2$ \\
  \hline
  $+_2$  & $r1$ \\ 
  \hline
  $\times$  & $r0$ \\ 
  \hline
\end{tabular}}
\end{subfigure} 
\begin{subfigure}{0.15\textwidth}
\centering
\scalebox{0.6}{
\begin{tabular}{cc}
  \hline
  Operation set \\
  \hline
  $x_1$  $x_2$ \\ 
  \hline
  $+_1$  $x_3$ $x_4$ \\ 
  \hline
  $+_2$  \\ 
  \hline
  $\times$ \\ 
  \hline
\end{tabular}}
\end{subfigure}
\begin{subfigure}{0.15\textwidth}
\centering
\scalebox{0.6}{
\begin{tabular}{c}
  \hline
  Operation set \\
  \hline
  $x_1$  $x_2$ \\ 
  \hline
  $+_1$  $x_3$ \\ 
  \hline
  $x_4$ \\
  \hline
  $+_2$ \\ 
  \hline
  $\times$ \\ 
  \hline
\end{tabular}}
\end{subfigure}
\begin{subfigure}{0.20\textwidth}
\scalebox{0.6}{%
\begin{tikzpicture}[xscale=0.7,yscale=0.5]
\pgfsetlinewidth{1pt}
\tikzstyle{vr}=[circle,minimum size=8mm,draw,fill=blue!40,drop shadow]
\tikzstyle{op}=[circle,minimum size=8mm,draw,fill=green!40,drop shadow]
\tikzstyle{ls}=[circle,minimum size=8mm,draw,fill=red!40,drop shadow]
\draw (0,2) node[vr] (x1) {$x_1$};
\draw (2,2) node[vr] (x2) {$x_2$};
\draw (4,2) node[vr] (x3) {$x_3$};
\draw (6,2) node[vr] (x4) {$x_4$};
\draw (1,4) node[op] (y1) {$+_1$};
\draw (5,4) node[op] (y2) {$+_2$};
\draw (2,6) node[ls] (st) {$st$};
\draw (3,8) node[ls] (ld) {$ld$};
\draw (4,10) node[op] (z1) {$\times$};
\draw[->](x1) to (y1);
\draw[->](x2) to (y1);
\draw[->](x3) to (y2);
\draw[->](x4) to (y2);
\draw[->](y1) to (st);
\draw[->](st) to (ld);
\draw[->](ld) to (z1);
\draw[->](y2) to (z1);
\end{tikzpicture}}
\end{subfigure}
\caption{(a) An example expression tree, (b) a depth-first ordering of operations and register assignment, (c) operations bundled together for concurrent execution, (d) operations bundled together for concurrent execution on a machine with only $2$ write ports in register file, and (e) an expression tree with spill code.}
\label{fig:riscexecute}
\end{figure}

As an example, consider the expression tree shown in Figure~\ref{fig:riscexecute}(a): With a register file containing $4$ registers and $4$ read/write ports, this simple expression can be evaluated in only $3$ steps if all levels are executed in parallel. With only $3$ registers, one can avoid using load/store instructions by ordering the tree nodes and assigning registers by a depth-first traversal in an optimal way according to \cite{SeUl70}. Figure~\ref{fig:riscexecute}(b) shows the operation order and the corresponding register assignment. The Figure~\ref{fig:riscexecute}(c) divides the sequence of operations into sets, where all operations in a set can be executed concurrently. Clearly, it takes $4$ steps to evaluate the expression tree in parallel this way. Note that even without the overhead of load and store instructions (i.e., given a sufficient number of registers to avoid memory accesses), it is apparent that the register-based code limits the use of ILP. Moreover, if only $2$ write ports are available in the register file, at least $5$ steps are then required to evaluate the expression as shown by the independent sets of operations in Figure~\ref{fig:riscexecute}(d). Note that at most $2$ operations can be bundled together since each operation must write its result to a register and only $2$ write ports are available in the register file. Finally, if only $2$ registers would be available, one has to insert spill code in that the obtained result of $x_1+x_2$ is temporarily stored in memory and loaded in a register after having evaluated $x_3+x_4$ as shown in Figure~\ref{fig:riscexecute}(e). The expression evaluation is then further delayed by the latency of the memory access. Hence, the number of programmer-accessible registers influences the use of ILP. Therefore, in a processor with many processing units for parallel execution, it is desirable to communicate values directly between processing units without using global registers in between.

\emph{Exposed datapath architectures} offer this feature by allowing the compiler to move values directly from one PU to another PU. These architectures usually contain a large number of PUs with local storage. However, their current compiler technology \cite{LBFS98,SBGM06,AJEK16,GLMM06,ScSL09}, although utilizing register bypassing, still relies on classic code generators where the use of a minimal number of registers is in the focus rather than maximizing the degree of ILP. In \cite{BhJS16}, we therefore suggested a code generation technique for exposed datapath architectures that is based on a breadth-first traversal rather than the classic depth-first traversal \cite{SeUl70} over the syntax trees. To that end, we considered the Synchronous Control Asynchronous Dataflow (SCAD) architecture that is a special exposed datapath architecture whose PUs have buffers for each input and output port. The classic depth-first traversal was motivated by the reuse of registers, while the breadth-first version was motivated by executing independent instructions in parallel. 

However, with a limited number of PUs, computational overhead in the form of swap and duplication operations becomes necessary for SCAD machines \cite{BhJS16} (see Section~\ref{chap:bkGround}). Whether an overhead-free schedule exists for a given program on a SCAD machine with a given number of PUs was formalized in \cite{BhSc16} as a satisfiability (SAT) problem. Repeated invocation of the SAT solver by incrementing the number of PUs allowed one to determine the minimal number of PUs required to execute programs without any computational overhead. Since execution time is a more important metric in instruction scheduling, we refined the SAT encoding in \cite{BhSc17} to a satisfiability modulo theories (SMT) problem, introducing execution time as an additional parameter for optimization. The non-linearity of buffer constraints (a property specific to SCAD as discussed in  Section~\ref{chap:optPU}) did not allow the use of integer linear programming solvers that is more commonly used in instruction scheduling. While the SAT solver could process up to $15$-instruction programs with a timeout of $5$ seconds, the SMT solver could only process up to $12$-instruction programs even with a timeout of $60$ seconds.  %\cite{LBFS98,SSMP07,BKMD04,TSBS07,WSCH15,Corp99,SHWC13} 

%-------------------------------------------------------------------------------
\subsection{Contributions}
%-------------------------------------------------------------------------------
In this paper, we refine the SAT encoding of \cite{BhSc16}) to an Answer Set Programming (ASP) problem and use Clingo \cite{GKNS07a} to obtain answer sets (schedules) of varying execution times. Execution time constraints can be intuitively included in the ASP encoding using a graph theoretic formulation. Paths of maximal length in the directed acyclic graph helps to find the schedule(s) with optimal execution time. Furthermore, the ASP encoding can be easily adapted to different optimization criteria like to minimize execution time on a given number of PUs, or to minimize the number of PUs to achieve a given execution time. The specific contributions in this paper are the following:

\begin{itemize}
\item We encode time constrained SCAD scheduling as an ASP problem by formulating straight-line programs (without branches) as directed acyclic graphs and determine optimal schedules using the answer set solver Clingo.
\item We show that the ASP encoding can be easily modified to solve optimization problems with different optimization criteria.
\item We provide experimental results that clearly demonstrate the improved feasibility of ASP-based SCAD code generation in processing larger programs compared to code generation using SAT and SMT solvers. 
\end{itemize}

%-------------------------------------------------------------------------------
\subsection{Outline}
%-------------------------------------------------------------------------------
The rest of the paper has the following outline: Section~\ref{chap:bkGround} presents sufficient details of the SCAD architecture and its code generation problem. Section~\ref{chap:optPU} considers then a first code optimization problem in terms of an ASP problem in that we determine the minimal number of PUs to generate overhead-free code (this was the SAT problem considered in \cite{BhSc16}). Section~\ref{chap:optTime} introduces then a new ASP formulation for determining the minimal execution time of a given straightline program and SCAD machine. Section~\ref{chap:optMore} generalizes both ASP problems in that we will then determine the the minimal number of PUs required to run a given program in a given bound of time or conversely the minimal execution time for a given number of PUs. Section~\ref{chap:benchmarks} presents experimental results that we obtained using the ASP solver Clingo. As we will show there, our new ASP formulation leads to much better results than the previous SAT and SMT encodings and is even more flexible in that we can consider different optimization problems in ASP alone. Finally, Section~\ref{chap:relWork} considers related work, and Section~\ref{chap:conclusion} lists our conclusions.

%% file: Chap2_Background.tex
%-------------------------------------------------------------------------------
\section{Background} \label{chap:bkGround}
%-------------------------------------------------------------------------------
In this section, we briefly describe the basics of SCAD architectures and the related code generation problem. For a more detailed description of SCAD and the optimal code generation problem discussed here see \cite{BhSc16}.

%-------------------------------------------------------------------------------
\subsection{SCAD Architectures}
\begin{figure}
\begin{subfigure}[b]{0.4\textwidth}
\begin{center}
\scalebox{1.0}[0.8]{
\includegraphics[scale=0.08]{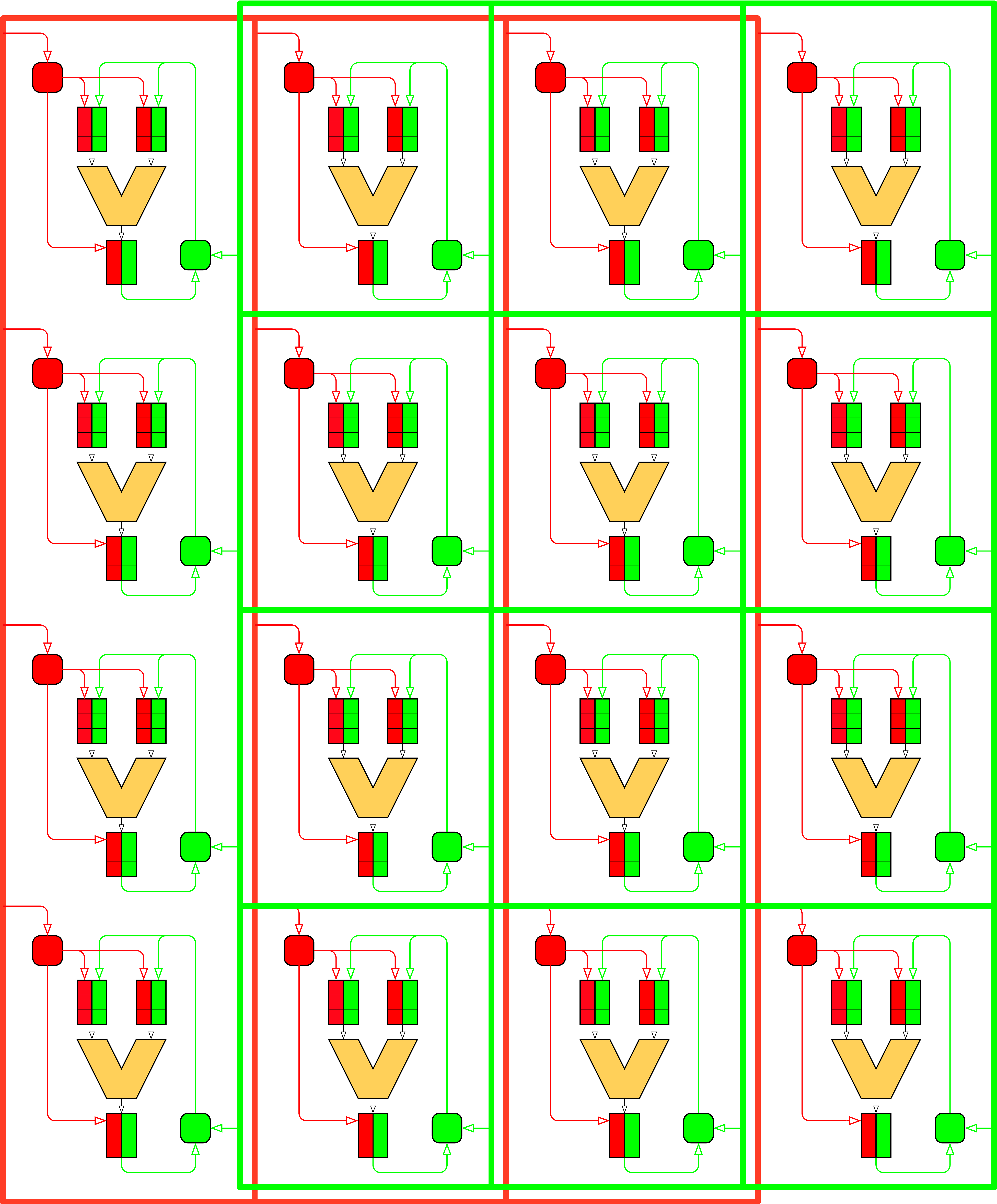}}
\end{center}
\end{subfigure}
\quad
\begin{subfigure}[b]{0.25\textwidth}
\includegraphics[scale=0.22]{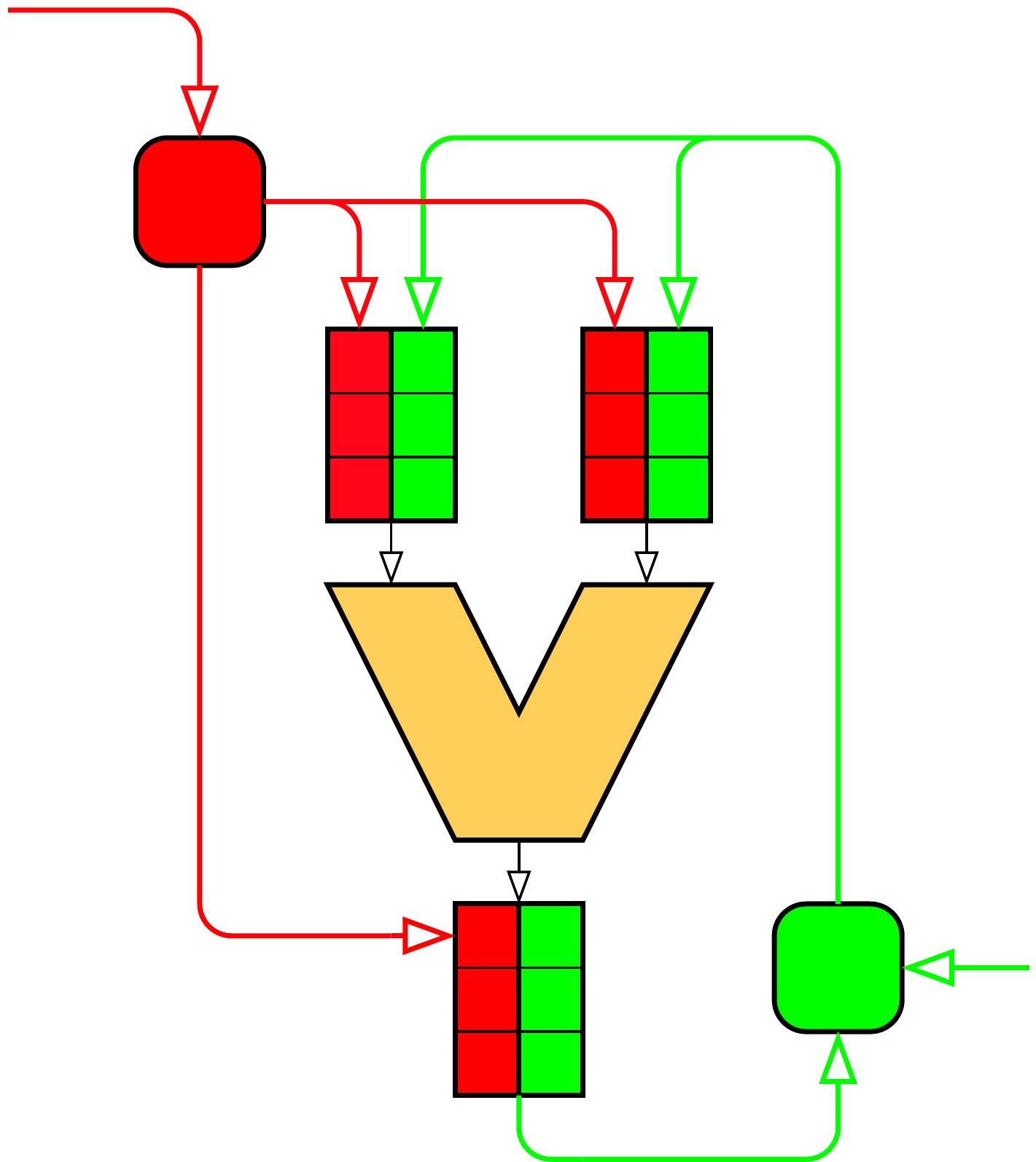}
\end{subfigure}
\quad
\quad
\begin{subfigure}[b]{0.25\textwidth}
\scalebox{0.275}{%
\includegraphics{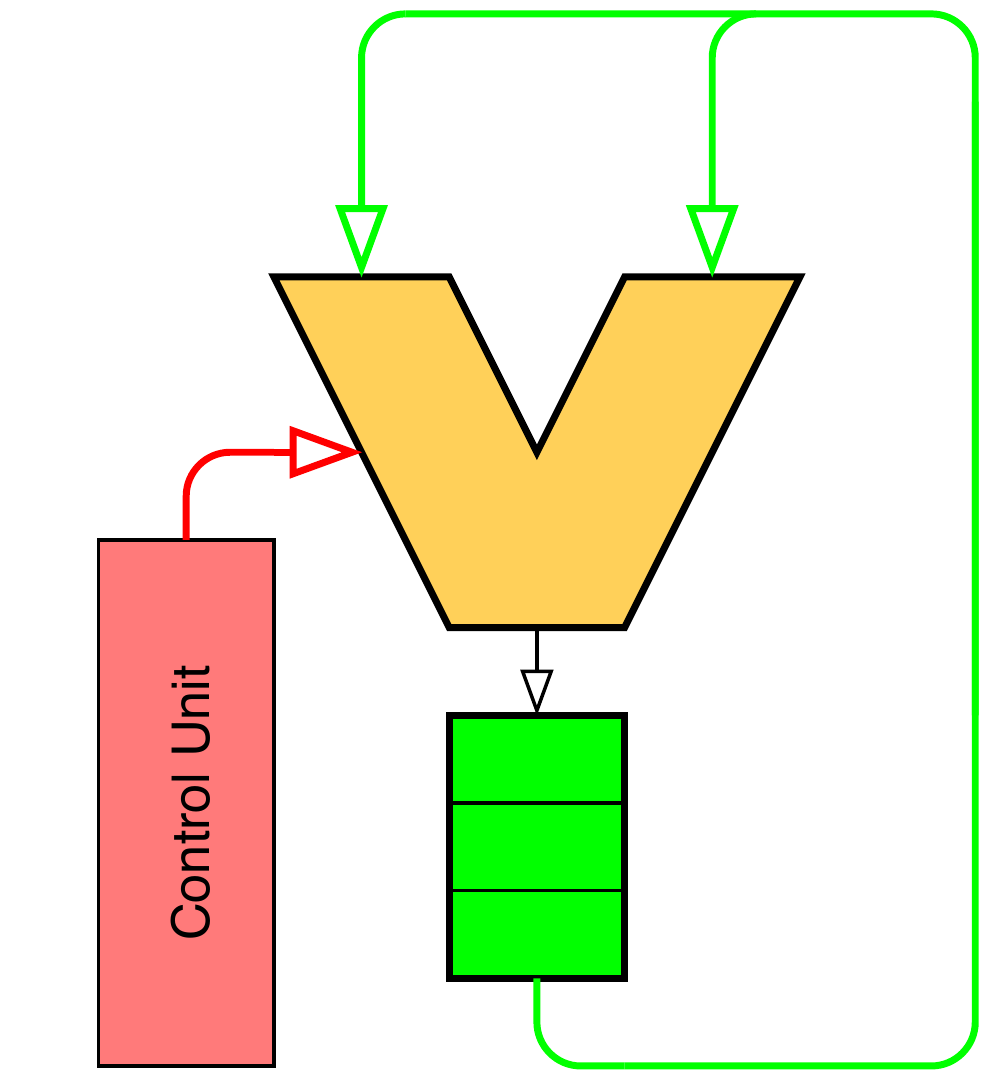}}
\end{subfigure}
\vspace{1ex}
\caption{(a) Architecture of a SCAD Processor (b) A processing unit in SCAD and (c) Architecture of queue machine}
\label{fig:scadQueue}
\end{figure}

%-------------------------------------------------------------------------------
The general organization of PUs in a SCAD architecture is shown in Figure~\ref{fig:scadQueue}(a). Although only two-input one-output PUs are shown, a PU in a SCAD architecture may also implement functions with an arbitrary number of inputs and outputs. Each PU shown in Figure~\ref{fig:scadQueue}(b) has queues (or first-in-first-out (FIFO) buffers) at its input and output ports. Input and output buffers have unique addresses which are used in move instructions which are the only kind of instructions of the SCAD machine: A move instruction $src\to tgt$ means that a value from output buffer $src$ shall be sent to input buffer $tgt$.

Input and output buffers are connected to two interconnection networks: The \emph{move-instruction bus (MIB)} (given in red color) is used to synchronously send move instructions from the control unit to the PUs owning output buffer $src$ and input buffer $tgt$. Second, the \emph{data transport network (DTN)} (given in green color) is used later by the PUs to asynchronously send values whenever these are available. All buffers store pairs $(adr,val)$ of entries. For an input buffer, $adr$ is the address of the output buffer of the PU that has already produced or that will produce the value $val$. An entry $(adr,\bot)$ with the special value $\bot$ is used to indicate that the required value is not yet available and will later be sent from the output buffer $adr$. Similarly, for an output buffer, $adr$ is the address of the input buffer of the PU that is waiting for the value $val$. An entry $(adr,\bot)$ with the special value $\bot$ is used to indicate that the required value is not yet available and will later be produced by the PU and can then be sent to the input buffer $adr$.

SCAD is programmed by a sequence of \emph{move instructions} $src\to tgt$ whose semantics is to move a value from the head of output buffer $src$ to the tail of input buffer $tgt$. The MIB is used to synchronously register move instructions $src\to tgt$ issued by the control unit on the PUs owning buffers $src$ and $tgt$, respectively. The PUs continuously snoop the MIB for move instructions $src\to tgt$. If an output buffer has address $src$, it will add the entry $(tgt,\bot)$ to its tail. At the same time, the input buffer with address $tgt$ will add the pair $(src,\bot)$ to its tail. If any of these buffers is full, a feedback signal $fullBuffer$ is given. Then, no buffer is written and the control unit (CU) is stalled until there is space in both buffers to register the move successfully. Therefore, the move instructions are registered synchronously. After this, the execution in the PUs and the actual data transports are carried out asynchronously once the values will be available.

Assume a PU with $n$ inputs and $m$ outputs has valid values $x_1,\ldots,x_n$ at the heads of its input buffers. If each of its $m$ output buffers has enough free space, the corresponding operation will fire. Thus, PU execution is only driven by the availability of input data. This will produce output values $f_1(x_1,\ldots, x_n)$, \ldots, $f_m(x_1,\ldots, x_n)$ for every output buffer. The pair $(adr,\bot)$ closest to the head of the $j^{th}$ output buffer is then replaced by $(adr, f_j(x_1,\ldots,x_n))$. 

A message in the DTN is a triple $(src,tgt,val)$ where $src$ is the output buffer where $val$ was produced and $tgt$ is the input buffer the message is being sent to. When an output buffer $src$ finds the pair $(tgt,val)$ with $val \neq \bot$ at its head, it will produce the message $(src,tgt,val)$ for transmission by DTN and will then remove the pair from the buffer. When receiving the message, the input buffer $tgt$ will replace the entry $(src,\bot)$ closest to its head with $(src,val)$.

There is at least one \emph{load/store unit} (LSU) that handles memory accesses that are of course also required for SCAD machines. As expected, the input buffers contain the address and value to be stored for a store operation, and the address and a target buffer's address for a load value.

Branch instructions are handled as follows by the CU: if the target of a move instruction is the CU itself, it is meant to be the program counter. In this case, the CU stops fetching move instructions and has to wait until this value arrives at the head of its input buffer associated with the program counter. Otherwise, it will simply increment the program counter and place it on its input buffer's head associated with the program counter so that the next move instruction in the program order is fetched in the subsequent clock cycle.

%-------------------------------------------------------------------------------
\subsection{Code Generation for SCAD}
%-------------------------------------------------------------------------------
A queue machine (see Figure~\ref{fig:scadQueue}(c)) \cite{FeEr81} reads operands for executing an operation from the head of a queue and adds the results to the tail of that queue. Generating a queue program to evaluate an expression tree is done by a breadth-first traversal of the tree \cite{FeEr81}. A consistent left to right or right to left traversal ensures that operands required to execute operations at one level are available in the queue in the correct order. 

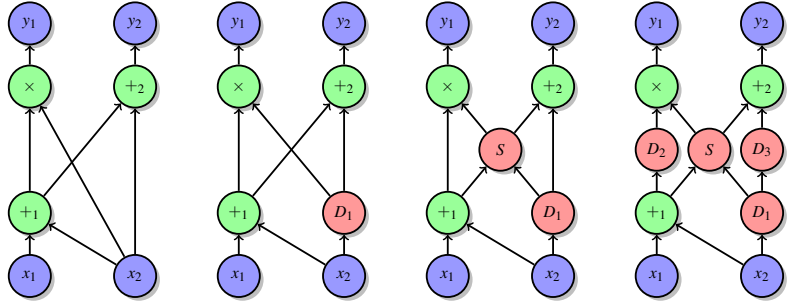
\begin{figure}[htb]
\begin{center}
% ------------------------------------------------------------------------------
% given expression dag
% ------------------------------------------------------------------------------
\begin{subfigure}{0.20\textwidth}
\scalebox{0.7}{
\begin{tikzpicture}[xscale=1.0,yscale=1.2]
\pgfsetlinewidth{1pt}
\tikzstyle{vr}=[circle,minimum size=8mm,draw,fill=blue!40,drop shadow]
\tikzstyle{op}=[circle,minimum size=8mm,draw,fill=green!40,drop shadow]
\draw (0,0) node[vr] (x1) {$x_1$};
\draw (2,0) node[vr] (x2) {$x_2$};
\draw (0,1) node[op] (o1) {$+_1$};
\draw (0,3) node[op] (o6) {$\times$};
\draw (2,3) node[op] (o7) {$+_2$};
\draw (0,4) node[vr] (y1) {$y_1$};
\draw (2,4) node[vr] (y2) {$y_2$};
\draw[->](x1) to (o1);
\draw[->](x2) to (o1);
\draw[->](x2) to (o6);
\draw[->](x2) to (o7);
\draw[->](o1) to (o6);
\draw[->](o1) to (o7);
\draw[->](o6) to (y1);
\draw[->](o7) to (y2);
\end{tikzpicture}}
\end{subfigure}
% ------------------------------------------------------------------------------
% levelized expression dag
% ------------------------------------------------------------------------------
\begin{subfigure}{0.20\textwidth}
\scalebox{0.7}{
\begin{tikzpicture}[xscale=1.0,yscale=1.2]
\pgfsetlinewidth{1pt}
\tikzstyle{vr}=[circle,minimum size=8mm,draw,fill=blue!40,drop shadow]
\tikzstyle{op}=[circle,minimum size=8mm,draw,fill=green!40,drop shadow]
\tikzstyle{sd}=[circle,minimum size=8mm,draw,fill=red!40,drop shadow]
\draw (0,0) node[vr] (x1) {$x_1$};
\draw (2,0) node[vr] (x2) {$x_2$};
\draw (0,1) node[op] (o1) {$+_1$};
\draw (2,1) node[sd] (o2) {$D_1$};
\draw (0,3) node[op] (o6) {$\times$};
\draw (2,3) node[op] (o7) {$+_2$};
\draw (0,4) node[vr] (y1) {$y_1$};
\draw (2,4) node[vr] (y2) {$y_2$};
\draw[->](x1) to (o1);
\draw[->](x2) to (o1);
\draw[->](x2) to (o2);
\draw[->](o1) to (o6);
\draw[->](o1) to (o7);
\draw[->](o2) to (o6);
\draw[->](o2) to (o7);
\draw[->](o6) to (y1);
\draw[->](o7) to (y2);
\end{tikzpicture}}
\end{subfigure}
% ------------------------------------------------------------------------------
% planarized expression dag
% ------------------------------------------------------------------------------
\begin{subfigure}{0.20\textwidth}
\scalebox{0.7}{
\begin{tikzpicture}[xscale=1.0,yscale=1.2]
\pgfsetlinewidth{1pt}
\tikzstyle{vr}=[circle,minimum size=8mm,draw,fill=blue!40,drop shadow]
\tikzstyle{op}=[circle,minimum size=8mm,draw,fill=green!40,drop shadow]
\tikzstyle{sd}=[circle,minimum size=8mm,draw,fill=red!40,drop shadow]
\draw (0,0) node[vr] (x1) {$x_1$};
\draw (2,0) node[vr] (x2) {$x_2$};
\draw (0,1) node[op] (o1) {$+_1$};
\draw (2,1) node[sd] (o2) {$D_1$};
\draw (1,2) node[sd] (o4) {$S$};
\draw (0,3) node[op] (o6) {$\times$};
\draw (2,3) node[op] (o7) {$+_2$};
\draw (0,4) node[vr] (y1) {$y_1$};
\draw (2,4) node[vr] (y2) {$y_2$};
\draw[->](x1) to (o1);
\draw[->](x2) to (o1);
\draw[->](x2) to (o2);
\draw[->](o1) to (o6);
\draw[->](o1) to (o4);
\draw[->](o2) to (o4);
\draw[->](o2) to (o7);
\draw[->](o4) to (o6);
\draw[->](o4) to (o7);
\draw[->](o6) to (y1);
\draw[->](o7) to (y2);
\end{tikzpicture}}
\end{subfigure}
% ------------------------------------------------------------------------------
% levelized and planarized expression dag
% ------------------------------------------------------------------------------
\begin{subfigure}{0.20\textwidth}
\scalebox{0.7}{
\begin{tikzpicture}[xscale=1.0,yscale=1.2]
\pgfsetlinewidth{1pt}
\tikzstyle{vr}=[circle,minimum size=8mm,draw,fill=blue!40,drop shadow]
\tikzstyle{op}=[circle,minimum size=8mm,draw,fill=green!40,drop shadow]
\tikzstyle{sd}=[circle,minimum size=8mm,draw,fill=red!40,drop shadow]
\draw (0,0) node[vr] (x1) {$x_1$};
\draw (2,0) node[vr] (x2) {$x_2$};
\draw (0,1) node[op] (o1) {$+_1$};
\draw (2,1) node[sd] (o2) {$D_1$};
\draw (0,2) node[sd] (o3) {$D_2$};
\draw (1,2) node[sd] (o4) {$S$};
\draw (2,2) node[sd] (o5) {$D_3$};
\draw (0,3) node[op] (o6) {$\times$};
\draw (2,3) node[op] (o7) {$+_2$};
\draw (0,4) node[vr] (y1) {$y_1$};
\draw (2,4) node[vr] (y2) {$y_2$};
\draw[->](x1) to (o1);
\draw[->](x2) to (o1);
\draw[->](x2) to (o2);
\draw[->](o1) to (o3);
\draw[->](o1) to (o4);
\draw[->](o2) to (o4);
\draw[->](o2) to (o5);
\draw[->](o3) to (o6);
\draw[->](o4) to (o6);
\draw[->](o4) to (o7);
\draw[->](o5) to (o7);
\draw[->](o6) to (y1);
\draw[->](o7) to (y2);
\end{tikzpicture}}
\end{subfigure}
\vspace{1ex}
\caption{(a) An expression DAG with its (b) levelized and (c) planarized version, and (d) the final level-planar expression DAG}
\label{fig:allDag}
\end{center}
\end{figure}

Basic blocks of programs are often represented as directed acyclic graphs (DAGs). Generating a queue program for an expression \emph{tree} is easy since an expression tree is by definition a level-planar graph \cite{ScLY02}. However, generating queue programs for general expression DAGs involves first converting the DAG into a \emph{level-planar graph} and then performing a breadth-first traversal of the graph \cite{ScLY02} as shown in Figure~\ref{fig:allDag}. The given expression DAG is first levelized which means that operations must only refer to operands at the same level. This can be easily achieved by introducing \textit{dup} operations which take a value from the head of the queue and add some number of copies of it to the tail of the queue. Then, the graph is planarized which means that crossing edges are removed by inserting \textit{swap} operations which take two values from the head of the queue, and add them in exchanged order to the tail of the queue. One can sometimes avoid the introduction of \textit{swap} operations by suitable ordering of input or output nodes, but not in general. Finally, another levelization is usually required since \textit{swap} operations may be placed at new levels. A consistent left to right or right to left traversal is now performed on the level-planar graph to generate the queue program.

The first approach to generate code for a given SCAD machine was to translate queue code to move code whose execution on the SCAD machine simulates the execution of the queue code on the queue machine \cite{BhJS16}. It was observed that SCAD requires less overhead compared to queue machines since SCAD machines have multiple buffers (associated with multiple PUs) while queue machines only have one central queue. In other words, any basic block can be executed on a SCAD machine without overhead provided there are enough PUs. The SAT encoding in \cite{BhSc16} was then used to determine the minimal number of PUs required in a SCAD machine to execute a given basic block without any overhead. This was refined to an SMT encoding in \cite{BhSc17} to determine the minimal execution time of a given basic block on a SCAD machine with a given number of PUs (greater than or equal to the minimal number required for overhead free execution).

%% file: Chap3_OptimalPUs.tex
\section{Optimal Scheduling Regarding Processing Units by ASP}
\label{chap:optPU}

The first question we want to answer by ASP is the one considered in \cite{BhSc16} using SAT solvers: Given a basic block $B$, determine the minimal number of PUs required to schedule $B$ on a SCAD machine without duplication and swap operations. Using ASP, we can additionally count the number of those schedules, and can further analyze their structure since we can compute \emph{all} optimal schedules.

\begin{figure}[htb]
\begin{subfigure}{.4\textwidth}
\scalebox{0.8}{
\begin{tikzpicture}[xscale=0.7,yscale=0.6]
\pgfsetlinewidth{1pt}
\tikzstyle{vr}=[circle,minimum size=8mm,draw,fill=blue!40,drop shadow]
\tikzstyle{op}=[circle,minimum size=8mm,draw,fill=green!40,drop shadow]
\draw (0,0) node[vr] (x0) {$x_0$};
\draw (2.25,0) node[vr] (x1) {$x_1$};
\draw (5.5,0) node[vr] (x2) {$x_2$};
\draw (2.75,2) node[op] (x3) {$x_3$};
\draw (4.5,2) node[op] (x4) {$x_4$};
\draw (0,4) node[op] (x5) {$x_5$};
\draw (1.5,4) node[op] (x6) {$x_6$};
\draw (5.5,4) node[op] (x7) {$x_7$};
\draw (7,4) node[op] (x8) {$x_8$};
\draw[->](x0) to (x3);
\draw[->](x0) to (x4);
\draw[->](x0) to (x5);
\draw[->](x1) to (x6);
\draw[->](x1) to (x3);
\draw[->](x2) to (x4);
\draw[->](x2) to (x7);
\draw[->](x2) to (x8);
\draw[->](x3) to (x5);
\draw[->](x3) to (x6);
\draw[->](x4) to (x7);
\draw[->](x4) to (x8);
\end{tikzpicture}}
\end{subfigure}
\begin{subfigure}{.4\textwidth}
\centering
\programmath
\[
\begin{array}{c}
operand(x_3, x_0, x_1).\\
operand(x_4, x_0, x_2).\\
operand(x_5, x_0, x_3).\\
operand(x_6, x_1, x_3).\\
operand(x_7, x_4, x_2).\\
operand(x_8, x_4, x_2).
\end{array}
\]
\unprogrammath
\end{subfigure}
\vspace{1ex}
\caption{(a) An example basic block as DAG, (b) the same DAG as ASP-code}
\label{fig:example_bb}
\end{figure}
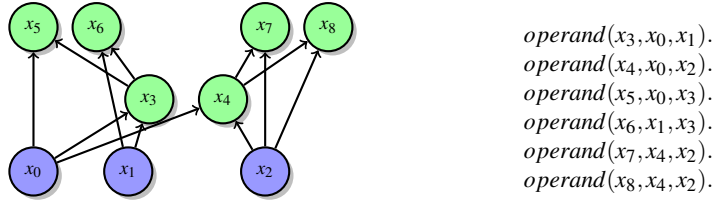

The input to our ASP encoding is a basic block that is represented as a list of variable assignments. The DAG representation and list of variable assignments for a basic block is shown in Figure~\ref{fig:example_bb}.

\noindent The expected output from the ASP encoding is/are schedule(s) of the given basic block on a SCAD machine. The schedule consists of two parts:
\begin{itemize}
\item \emph{Assignment:} which variable is produced by which PU?
\item \emph{Ordering:} in which order are the variables assigned to the same PU produced?
\end{itemize}
\noindent Assignment or mapping of variables to PUs and the ordering of variables on PUs must allow generation of move code to execute the basic block on the SCAD machine without any duplication and swap overhead. In other words, with the given variable assignment and ordering, one must be able to move values from output buffers to input buffers of PUs such that firing of computations on PUs produces values in the given order. Recall that all buffers are FIFO buffers and a PU fires if values are available at the heads of its input buffers. For example, a few valid schedules of the basic block in \autoref{fig:example_bb} are shown in \autoref{fig:example_bb_solutions_pu_min}. Note that although only one copy of each variable is shown, during actual execution as many copies of a variable are produced in the output buffer as needed in the basic block. For Solution $1$, the following is a valid sequence of move instructions that executes the basic block on the SCAD machine respecting the given order of variables (every $PU_n$ has two input buffers, the left one named $PU_{n,L}$ and the right one $PU_{n,R}$): 
$x_0 \rightarrow PU_{0, L}$, 
$x_0 \rightarrow PU_{0, L}$, 
$x_0 \rightarrow PU_{1, L}$, 
$x_1 \rightarrow PU_{0, L}$, 
$x_1 \rightarrow PU_{1, R}$, 
$x_2 \rightarrow PU_{1, R}$, 
$x_2 \rightarrow PU_{1, R}$, 
$x_2 \rightarrow PU_{0, R}$, 
$x_3 \rightarrow PU_{0, R}$, 
$x_3 \rightarrow PU_{0, R}$, 
$x_4 \rightarrow PU_{1, L}$,
$x_4 \rightarrow PU_{1, L}$.
Similarly, one can generate valid move code for other solutions. It should also be noted that values of the variables $x_0$, $x_1$ and $x_2$ are not computed, but loaded from the main memory. To load a variable value, we simply move the memory address of that variable to the left input buffer of the PU to which the variable is assigned. Since the memory addresses are known beforehand, we may place this move anywhere in the above sequence of moves and are not shown for simplicity. 

\begin{figure}[htb]
\begin{tabular}{|c|c|c|}
\hline\\
\begin{subfigure}[t]{0.3\textwidth}
%Solution 2, PU0
\begin{subfigure}[t]{0.475\textwidth}
\begin{tikzpicture}[scale=0.3,baseline=32]
\draw  (0.5,3) rectangle (-1,4.5);
\node at (-0.25,3.75) {};
\draw  (-3,3) rectangle (-4.5,4.5);
\node at (-3.75,3.75) {};

\draw [fill=orange] (-2.75,-1) -- (-1.25,-1) -- (0.5,2) -- (-1,2) -- (-2,0.5) -- (-3,2) -- (-4.5,2) -- cycle;
\node at (-2,-0.25) {$PU_0$};

\draw  (-2.75,-2) rectangle (-1.25,-3.5);
\node at (-2,-2.75) {$x_6$};

\draw  (-2.75,-3.5) rectangle (-1.25,-5);
\node at (-2,-4.25) {$x_5$};

\draw  (-2.75,-5) rectangle (-1.25,-6.5);
\node at (-2,-5.75) {$x_4$};

\draw  (-2.75,-6.5) rectangle (-1.25,-8);
\node at (-2,-7.25) {$x_1$};

\draw  (-2.75,-8) rectangle (-1.25,-9.5);
\node at (-2,-8.75) {$x_0$};

\draw [->] (-3.75,3) -- (-3.75,2);
\draw [->] (-0.25,3) -- (-0.25,2);
\draw [->] (-2,-1) -- (-2,-2);
\end{tikzpicture}
\end{subfigure}
%Solution 2, PU1
\begin{subfigure}[t]{0.475\textwidth}
\begin{tikzpicture}[scale=0.3,baseline=32]
\draw  (0.5,3) rectangle (-1,4.5);
\node at (-0.25,3.75) {};
\draw  (-3,3) rectangle (-4.5,4.5);
\node at (-3.75,3.75) {};

\draw [fill=orange] (-2.75,-1) -- (-1.25,-1) -- (0.5,2) -- (-1,2) -- (-2,0.5) -- (-3,2) -- (-4.5,2) -- cycle;
\node at (-2,-0.25) {$PU_1$};

\draw  (-2.75,-2) rectangle (-1.25,-3.5);
\node at (-2,-2.75) {$x_8$};

\draw  (-2.75,-3.5) rectangle (-1.25,-5);
\node at (-2,-4.25) {$x_7$};

\draw  (-2.75,-5) rectangle (-1.25,-6.5);
\node at (-2,-5.75) {$x_3$};

\draw  (-2.75,-6.5) rectangle (-1.25,-8);
\node at (-2,-7.25) {$x_2$};

\draw [->] (-3.75,3) -- (-3.75,2);
\draw [->] (-0.25,3) -- (-0.25,2);
\draw [->] (-2,-1) -- (-2,-2);
\end{tikzpicture}
\end{subfigure}
\caption{Solution 1}
\end{subfigure}
&
%Solution 3, PU0
\begin{subfigure}[t]{0.3\textwidth}
\begin{subfigure}[t]{0.475\textwidth}
\begin{tikzpicture}[scale=0.3,baseline=32]
\draw  (0.5,3) rectangle (-1,4.5);
\node at (-0.25,3.75) {};
\draw  (-3,3) rectangle (-4.5,4.5);
\node at (-3.75,3.75) {};

\draw [fill=orange] (-2.75,-1) -- (-1.25,-1) -- (0.5,2) -- (-1,2) -- (-2,0.5) -- (-3,2) -- (-4.5,2) -- cycle;
\node at (-2,-0.25) {$PU_0$};

\draw  (-2.75,-2) rectangle (-1.25,-3.5);
\node at (-2,-2.75) {$x_8$};

\draw  (-2.75,-3.5) rectangle (-1.25,-5);
\node at (-2,-4.25) {$x_4$};

\draw  (-2.75,-5) rectangle (-1.25,-6.5);
\node at (-2,-5.75) {$x_5$};

\draw  (-2.75,-6.5) rectangle (-1.25,-8);
\node at (-2,-7.25) {$x_1$};

\draw  (-2.75,-8) rectangle (-1.25,-9.5);
\node at (-2,-8.75) {$x_0$};

\draw [->] (-3.75,3) -- (-3.75,2);
\draw [->] (-0.25,3) -- (-0.25,2);
\draw [->] (-2,-1) -- (-2,-2);
\end{tikzpicture}
\end{subfigure}
%Solution 3, PU1
\begin{subfigure}[t]{0.475\textwidth}
\begin{tikzpicture}[scale=0.3,baseline=32]
\draw  (0.5,3) rectangle (-1,4.5);
\node at (-0.25,3.75) {};
\draw  (-3,3) rectangle (-4.5,4.5);
\node at (-3.75,3.75) {};

\draw [fill=orange] (-2.75,-1) -- (-1.25,-1) -- (0.5,2) -- (-1,2) -- (-2,0.5) -- (-3,2) -- (-4.5,2) -- cycle;
\node at (-2,-0.25) {$PU_1$};

\draw  (-2.75,-2) rectangle (-1.25,-3.5);
\node at (-2,-2.75) {$x_7$};

\draw  (-2.75,-3.5) rectangle (-1.25,-5);
\node at (-2,-4.25) {$x_6$};

\draw  (-2.75,-5) rectangle (-1.25,-6.5);
\node at (-2,-5.75) {$x_2$};

\draw  (-2.75,-6.5) rectangle (-1.25,-8);
\node at (-2,-7.25) {$x_3$};

\draw [->] (-3.75,3) -- (-3.75,2);
\draw [->] (-0.25,3) -- (-0.25,2);
\draw [->] (-2,-1) -- (-2,-2);
\end{tikzpicture}
\end{subfigure}
\caption{Solution 2}
\end{subfigure}%\\
&
\begin{subfigure}[t]{0.3\textwidth}
%Solution 2 (sym), PU0
\begin{subfigure}[t]{0.475\textwidth}
\begin{tikzpicture}[scale=0.3,baseline=32]
\draw  (0.5,3) rectangle (-1,4.5);
\node at (-0.25,3.75) {};
\draw  (-3,3) rectangle (-4.5,4.5);
\node at (-3.75,3.75) {};

\draw [fill=orange] (-2.75,-1) -- (-1.25,-1) -- (0.5,2) -- (-1,2) -- (-2,0.5) -- (-3,2) -- (-4.5,2) -- cycle;
\node at (-2,-0.25) {$PU_0$};

\draw  (-2.75,-2) rectangle (-1.25,-3.5);
\node at (-2,-2.75) {$x_8$};

\draw  (-2.75,-3.5) rectangle (-1.25,-5);
\node at (-2,-4.25) {$x_7$};

\draw  (-2.75,-5) rectangle (-1.25,-6.5);
\node at (-2,-5.75) {$x_3$};

\draw  (-2.75,-6.5) rectangle (-1.25,-8);
\node at (-2,-7.25) {$x_2$};

\draw [->] (-3.75,3) -- (-3.75,2);
\draw [->] (-0.25,3) -- (-0.25,2);
\draw [->] (-2,-1) -- (-2,-2);
\end{tikzpicture}
\end{subfigure}
%Solution 2 (sym), PU1
\begin{subfigure}[t]{0.475\textwidth}
\begin{tikzpicture}[scale=0.3,baseline=32]
\draw  (0.5,3) rectangle (-1,4.5);
\node at (-0.25,3.75) {};
\draw  (-3,3) rectangle (-4.5,4.5);
\node at (-3.75,3.75) {};

\draw [fill=orange] (-2.75,-1) -- (-1.25,-1) -- (0.5,2) -- (-1,2) -- (-2,0.5) -- (-3,2) -- (-4.5,2) -- cycle;
\node at (-2,-0.25) {$PU_1$};

\draw  (-2.75,-2) rectangle (-1.25,-3.5);
\node at (-2,-2.75) {$x_6$};

\draw  (-2.75,-3.5) rectangle (-1.25,-5);
\node at (-2,-4.25) {$x_5$};

\draw  (-2.75,-5) rectangle (-1.25,-6.5);
\node at (-2,-5.75) {$x_4$};

\draw  (-2.75,-6.5) rectangle (-1.25,-8);
\node at (-2,-7.25) {$x_1$};

\draw  (-2.75,-8) rectangle (-1.25,-9.5);
\node at (-2,-8.75) {$x_0$};

\draw [->] (-3.75,3) -- (-3.75,2);
\draw [->] (-0.25,3) -- (-0.25,2);
\draw [->] (-2,-1) -- (-2,-2);
\end{tikzpicture}
\end{subfigure}
\caption{Solution 1 (sym)}
\end{subfigure}\\
\hline
\end{tabular}
\caption{Three valid variable assignments and orderings that allow computation of the basic block in \autoref{fig:example_bb} without any overhead}
\label{fig:example_bb_solutions_pu_min}
\end{figure}

In our ASP encoding, we start by extracting all variables from our basic block/program with the following code.
\programmath
\[
\begin{array}{rcl}
    var(X) &:-& operand(X,Y,Z).\\
    var(Y) &:-& operand(X,Y,Z).\\
    var(Z) &:-& operand(X,Y,Z).
\end{array}
\]
\unprogrammath
\noindent Our generate part demands every variable to be assigned to exactly one PU, in the sense that the variable should be produced by this PU and put to the output buffer. Furthermore, if two values $V_1$ and $V_2$ are assigned to the same PU, they have to be ordered (either $V_1<V_2$ or $V_2<V_1$).
\programmath
\[
\begin{array}{rcl}
1\ \{asgn(VAR, PU) : pu(PU)\}\ 1&:-&var(VAR).\\
order(V_1,\ V_2),\ order(V_2, V_1) &:-& asgn(V_1,\ PU),\ asgn(V_2,\ PU),\ V_1\neq V_2.
\end{array}
\]
\unprogrammath
\noindent Our test part demands the order to fulfill additional constraints to allow the variable productions in a way that the computations of the basic block still can be performed. As shown in \cite{BhSc16} and \cite{BhSc17}, it is sufficient that the order of two variables $V_1,V_2$ produced by a PU $PU$ is preserved in their operands $V_{1,L/R}$ and $V_{2,L/R}$, if both operands are produced by the same PU $PU_2$.
\programmath
\[
\begin{array}{rcl}
order(V_{1,L},\ V_{2,L}) &:-& asgn(V_1,\ PU),\ asgn(V_2,\ PU),\ V_1\neq V_2,\ order(V_1,\  V_2),\\&&
                    operand(V_1,\ V_{1,L},\ V_{1,R}),\ operand(V_2,\ V_{2,L},\ V_{2,R}),\\&& asgn(V_{1,L},\ PU_2),\ asgn(V_{2,L},\ PU_2),\ V_{1,L}\neq V_{2,L}.\\
    order(V_{1,R},\ V_{2,R})&:-&asgn(V_1,\ PU),\ asgn(V_2,\ PU),\ V_1\neq V_2,\ order(V_1,\ V_2),\\&& operand(V_1,\ V_{1,L},\ V_{1,R}),\ operand(V_2,\ V_{2,L},\ V_{2,R}),\\&& asgn(V_{1,R},\ PU_2),\ asgn(V_{2,R},\ PU_2),\ V_{1,R}\neq V_{2,R}.
\end{array}
\]
\unprogrammath
\noindent Additionally, the order must also preserve the data dependencies of the basic block/DAG. For example, if an operand $V_{1.L}$ is produced by the same PU as its consumer $V_{1}$, it is required that first the operand is produced and afterwards its consumer. This must be generalized not only to the operands, but to all predecessors of a variable $V_1$ in the initial DAG. Therefore, the initial DAG is constructed, and a predicate that holds if a variable is a predecessor in this DAG. This predicate can then be used in the constraint to remove all orders which do not correspond to the predecessors order.

\programmath
\[
\begin{array}{rcl}
node(X) &:-& var(X).\\
    edge\_initial(X,Y) &:-& operand(Y,X,\_).\\
    edge\_initial(X,Y)& :- &operand(Y,\_,X).\\
    rootNode(X)& :-& not\ edge\_initial(\_,X),\ node(X).\\
    \\
    predecessor(X,Y) &:-& edge\_initial(X,Y).\\
    predecessor(X,Z) &:-& predecessor(X,Y),~predecessor(Y,Z).\\
  \\  
&:-& predecessor(X,Y),~asgn(X,~PU),~asgn(Y,~PU),~order(Y,X).
\end{array}
\]
\unprogrammath
\noindent To conclude, the transitive closure of the ordering relation is used in order to be able to compare every position of every variable on the same PU.
\programmath
\[
\begin{array}{rcl}
order(V_1,\ V_3)&:-&order(V_1, V_2),\ order(V_2,\ V_3).
\end{array}
\]
\unprogrammath
\noindent Using this basic encoding, we can now use ASP to find the minimal amount of PUs needed to allow the execution of a given basic block:
\programmath
\[
\begin{array}{l}
    pu(0..PUS-1)\ :-\ amountPUs(PUS).\\
    possiblePUAmount(0..max\_pus).\\
    1\ \{amountPUs(N):\ possiblePUAmount(N)\}\ 1.\\
    \#minimize\{N:amountPUs(N)\}.\\
\end{array}
\]
\unprogrammath
\noindent The above encoding led to 8800 different solutions, 3 of them have been picked out and are shown in \autoref{fig:example_bb_solutions_pu_min}. At a closer look, one can realize that the example (c) is a symmetric solution of (a). To avoid this overhead, symmetry breaking constraints have been added, which basically sort the PUs by the minimally assigned variable.
\programmath
\[
\begin{array}{rcl}
    minimum(PU, S) &:-& S=\#min\{VAR: asgn(VAR, PU)\}, pu(PU), asgn(\_, PU).\\
    &:-& minimum(PU,S), minimum(PU_2, S_2), PU_2>PU, S>S_2.
\end{array}
\]
\unprogrammath
\noindent This symmetry breaking constraint removes all rotations and combinations of PUs with the corresponding assignments. In numbers, $\frac{N}{amountPU!}$ many solutions are left if there are $N$ schedules given for $amountPU$ many PUs. For example, if there are $N=2016$ schedules of basic block on 4 PUs, $\frac{2016}{24} = 84$ unique solutions are left  after symmetry breaking. In our example (see \autoref{fig:example_bb}), the amount of results given by ASP is reduced by $\frac{8800}{2!}$ many solutions from 8800 to 4400.

%% file: Chap4_OptimalTime.tex
\section{Optimal Scheduling Regarding Execution Time}
\label{chap:optTime}

The main contribution of this paper is a new method to determine the execution time for scheduling a basic block. Similar to the SMT encoding \cite{BhSc17}, an assignment of variables to timeslots is used in this encoding. However, this approach did not scale well as the number of constraints grows enormously with the number of variables. Using ASP, we found a more direct way for determining the minimal execution time.

As stated in the previous section, the initial program/basic block can be represented by a DAG. The basic information added by a schedule is the introduction of additional edges into this DAG (all orders from the same PU which have not been ordered in the initial DAG). All answer sets are therefore the same as introducing all possibilities of additional edges into the graph with the only constraint that there are only minimal possible paths through the graph (minimal amount of PUs) without reaching a cycle.

From an abstract perspective, the longest path though this non-cyclic combined graph is the execution time/execution steps needed to compute the solutions of the basic block. By adding different costs for every OP-code, PU or adding costs for moving values, this can be easily refined to a particular real hardware implementation.

We illustrate our new approach by an example: Consider solution a) from \autoref{fig:example_bb_solutions_pu_min}. It has the additional edges $x_0 \rightarrow x_1$, $x_1 \rightarrow x_4$, $x_4 \rightarrow x_5$, and $x_5 \rightarrow x_6$ introduced by the order of $PU_0$ and $x_2 \rightarrow x_3$, $x_3 \rightarrow x_7$, and $x_7 \rightarrow x_8$ introduced by the order of $PU_1$. If we add these edges to the initial DAG, we can find a longest path through this combined DAG representing the amount of steps needed to execute the basic block with this order on a SCAD machine. In case of solution 1, there are at most 5 steps.

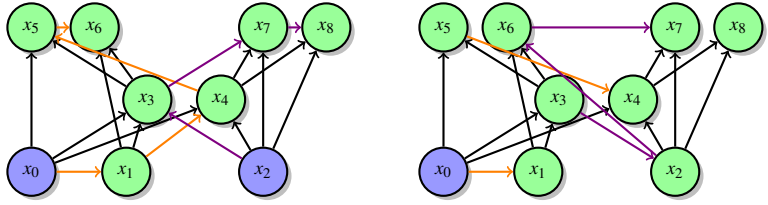
\begin{figure}[htb]
\begin{subfigure}{.4\textwidth}
\scalebox{0.8}{
\begin{tikzpicture}[xscale=0.7,yscale=0.6]
\pgfsetlinewidth{1pt}
\tikzstyle{vr}=[circle,minimum size=8mm,draw,fill=blue!40,drop shadow]
\tikzstyle{op}=[circle,minimum size=8mm,draw,fill=green!40,drop shadow]
\draw (0,0) node[vr] (x0) {$x_0$};
\draw (2.25,0) node[op] (x1) {$x_1$};
\draw (5.5,0) node[vr] (x2) {$x_2$};
\draw (2.75,2) node[op] (x3) {$x_3$};
\draw (4.5,2) node[op] (x4) {$x_4$};
\draw (0,4) node[op] (x5) {$x_5$};
\draw (1.5,4) node[op] (x6) {$x_6$};
\draw (5.5,4) node[op] (x7) {$x_7$};
\draw (7,4) node[op] (x8) {$x_8$};
\draw[->](x0) to (x3);
\draw[->](x0) to (x4);
\draw[->](x0) to (x5);
\draw[->](x1) to (x6);
\draw[->](x1) to (x3);
\draw[->](x2) to (x4);
\draw[->](x2) to (x7);
\draw[->](x2) to (x8);
\draw[->](x3) to (x5);
\draw[->](x3) to (x6);
\draw[->](x4) to (x7);
\draw[->](x4) to (x8);
%% additional edges from the result PU0
\draw[orange][->](x0) to (x1);
\draw[orange][->](x1) to (x4);
\draw[orange][->](x4) to (x5);
\draw[orange][->](x5) to (x6);
%% additional edges from the result PU1
\draw[violet][->](x2) to (x3);
\draw[violet][->](x3) to (x7);
\draw[violet][->](x7) to (x8);
\end{tikzpicture}}
\end{subfigure}
\begin{subfigure}{.4\textwidth}
\scalebox{0.8}{
\begin{tikzpicture}[xscale=0.7,yscale=0.6]
\pgfsetlinewidth{1pt}
\tikzstyle{vr}=[circle,minimum size=8mm,draw,fill=blue!40,drop shadow]
\tikzstyle{op}=[circle,minimum size=8mm,draw,fill=green!40,drop shadow]
\draw (0,0) node[vr] (x0) {$x_0$};
\draw (2.25,0) node[op] (x1) {$x_1$};
\draw (5.5,0) node[op] (x2) {$x_2$};
\draw (2.75,2) node[op] (x3) {$x_3$};
\draw (4.5,2) node[op] (x4) {$x_4$};
\draw (0,4) node[op] (x5) {$x_5$};
\draw (1.5,4) node[op] (x6) {$x_6$};
\draw (5.5,4) node[op] (x7) {$x_7$};
\draw (7,4) node[op] (x8) {$x_8$};
\draw[->](x0) to (x3);
\draw[->](x0) to (x4);
\draw[->](x0) to (x5);
\draw[->](x1) to (x6);
\draw[->](x1) to (x3);
\draw[->](x2) to (x4);
\draw[->](x2) to (x7);
\draw[->](x2) to (x8);
\draw[->](x3) to (x5);
\draw[->](x3) to (x6);
\draw[->](x4) to (x7);
\draw[->](x4) to (x8);
%% additional edges from the result PU0
\draw[orange][->](x0) to (x1);
\draw[orange][->](x5) to (x4);
%% additional edges from the result PU1
\draw[violet][->](x3) to (x2);
\draw[violet][->](x2) to (x6);
\draw[violet][->](x6) to (x7);
\end{tikzpicture}}
\end{subfigure}
\vspace{1ex}
\caption{Combined DAG for Solutions 1 and 2 from \autoref{fig:example_bb_solutions_pu_min}(a) and (b)}
\label{fig:example_bb_combined_dag}
\end{figure}

However, if we look at the combined DAG for Solution 2, we can see that there exists at least one path with cost 6. Although Solution 1 and 2 can both be scheduled on 2 PUs, Solution 1 would be a better choice for an optimal compiler as it requires less steps. This minimization can also be modeled in the ASP encoding: First, we constructed the combined DAG by adding all order relations introduced by our result to the initial DAG.
\programmath
\[
\begin{array}{rcl}
    edge(X,Y) &:-& edge\_initial(X,Y).\\
    edge(X,Y) &:-& order(X,Y).\\
    initialNode(X)&:-& not\ edge(\_,X),\ node(X).
\end{array}
\]
\unprogrammath
\noindent With the combined DAG, which represents all variable dependencies, we can calculate the longest path and take this as the execution time. Here, it is assumed that the combined DAG has no cycles, and therefore the length of the path can be used as cost measure.
\programmath
\[
\begin{array}{rcl}
pathCosts(X, 1) &:-& initialNode(X).\\
pathCosts(Y, N+1)& :-& edge(X,Y),\ pathCosts(X,N),\ N<(M+1),\ amountVars(M).\\
\\
maximalCost(N)& :-& N=\#max\{C: pathCosts(\_,C)\}.
\end{array}
\]
\unprogrammath
Finally, it is possible to answer the minimization question with a final minimize statement.
\programmath
\[
\begin{array}{c}
\#minimize\{N@1:\  maximalCost(N), N>\#inf\}.\\
    :- maximalCost(\#inf).
\end{array}
\]
\unprogrammath
With this minimize statement, we can determine all solutions that have minimal execution time for a given minimal number of PUs (the infimum $\#inf$ had to be used to ensure that the minimize statement does not allow to have no $maximalCost$ defined at all). As our total results for \autoref{fig:example_bb} to be scheduled on 2 PUs have been 8800 possible schedules, it was figured out that only 520 of them are optimal (260 with symmetry breaking). The solutions (a) and (c) shown in \autoref{fig:example_bb_solutions_pu_min} are minimal and have cost 5. Solution (b) is not optimal as it has cost 6.

However, are there even better schedules possible on a SCAD machine with more PUs? And how many PUs are needed to execute a program/basic block as fast as possible (with minimum execution time at all)? The answer to the first question is simple: The depth of the initial DAG represents the overall lower bound to the execution time. Therefore, the more interesting question is the second one.

\begin{figure}[htb]
%Solution PU0
\begin{subfigure}[t]{0.24\textwidth}
%Solution PU0
\begin{tikzpicture}[scale=0.3,baseline=32]
\draw  (0.5,3) rectangle (-1,4.5);
\node at (-0.25,3.75) {};
\draw  (-3,3) rectangle (-4.5,4.5);
\node at (-3.75,3.75) {};

\draw [fill=orange] (-2.75,-1) -- (-1.25,-1) -- (0.5,2) -- (-1,2) -- (-2,0.5) -- (-3,2) -- (-4.5,2) -- cycle;
\node at (-2,-0.25) {$PU_0$};

\draw  (-2.75,-2) rectangle (-1.25,-3.5);
\node at (-2,-2.75) {$x_8$};

\draw  (-2.75,-3.5) rectangle (-1.25,-5);
\node at (-2,-4.25) {$x_0$};

\draw [->] (-3.75,3) -- (-3.75,2);
\draw [->] (-0.25,3) -- (-0.25,2);
\draw [->] (-2,-1) -- (-2,-2);
\end{tikzpicture}
\end{subfigure}
%Solution PU1
\begin{subfigure}[t]{0.24\textwidth}
\begin{tikzpicture}[scale=0.3,baseline=32]
\draw  (0.5,3) rectangle (-1,4.5);
\node at (-0.25,3.75) {};
\draw  (-3,3) rectangle (-4.5,4.5);
\node at (-3.75,3.75) {};

\draw [fill=orange] (-2.75,-1) -- (-1.25,-1) -- (0.5,2) -- (-1,2) -- (-2,0.5) -- (-3,2) -- (-4.5,2) -- cycle;
\node at (-2,-0.25) {$PU_1$};

\draw  (-2.75,-2) rectangle (-1.25,-3.5);
\node at (-2,-2.75) {$x_6$};

\draw  (-2.75,-3.5) rectangle (-1.25,-5);
\node at (-2,-4.25) {$x_3$};

\draw  (-2.75,-5) rectangle (-1.25,-6.5);
\node at (-2,-5.75) {$x_1$};

\draw [->] (-3.75,3) -- (-3.75,2);
\draw [->] (-0.25,3) -- (-0.25,2);
\draw [->] (-2,-1) -- (-2,-2);
\end{tikzpicture}
\end{subfigure}
%Solution PU2
\begin{subfigure}[t]{0.24\textwidth}
\begin{tikzpicture}[scale=0.3,baseline=32]
\draw  (0.5,3) rectangle (-1,4.5);
\node at (-0.25,3.75) {};
\draw  (-3,3) rectangle (-4.5,4.5);
\node at (-3.75,3.75) {};

\draw [fill=orange] (-2.75,-1) -- (-1.25,-1) -- (0.5,2) -- (-1,2) -- (-2,0.5) -- (-3,2) -- (-4.5,2) -- cycle;
\node at (-2,-0.25) {$PU_2$};

\draw  (-2.75,-2) rectangle (-1.25,-3.5);
\node at (-2,-2.75) {$x_5$};

\draw  (-2.75,-3.5) rectangle (-1.25,-5);
\node at (-2,-4.25) {$x_2$};

\draw [->] (-3.75,3) -- (-3.75,2);
\draw [->] (-0.25,3) -- (-0.25,2);
\draw [->] (-2,-1) -- (-2,-2);
\end{tikzpicture}
\end{subfigure}
%Solution PU3
\begin{subfigure}[t]{0.24\textwidth}
\begin{tikzpicture}[scale=0.3,baseline=32]
\draw  (0.5,3) rectangle (-1,4.5);
\node at (-0.25,3.75) {};
\draw  (-3,3) rectangle (-4.5,4.5);
\node at (-3.75,3.75) {};

\draw [fill=orange] (-2.75,-1) -- (-1.25,-1) -- (0.5,2) -- (-1,2) -- (-2,0.5) -- (-3,2) -- (-4.5,2) -- cycle;
\node at (-2,-0.25) {$PU_3$};

\draw  (-2.75,-2) rectangle (-1.25,-3.5);
\node at (-2,-2.75) {$x_7$};

\draw  (-2.75,-3.5) rectangle (-1.25,-5);
\node at (-2,-4.25) {$x_4$};

\draw [->] (-3.75,3) -- (-3.75,2);
\draw [->] (-0.25,3) -- (-0.25,2);
\draw [->] (-2,-1) -- (-2,-2);
\end{tikzpicture}
\end{subfigure}
\vspace{1ex}
\caption{One example to minimize the amount of execution time and then find the minimal number of PUs needed to schedule the example basic block shown in \autoref{fig:example_bb}}
\label{fig:example_bb_solutions_exec_minimization}
\end{figure}
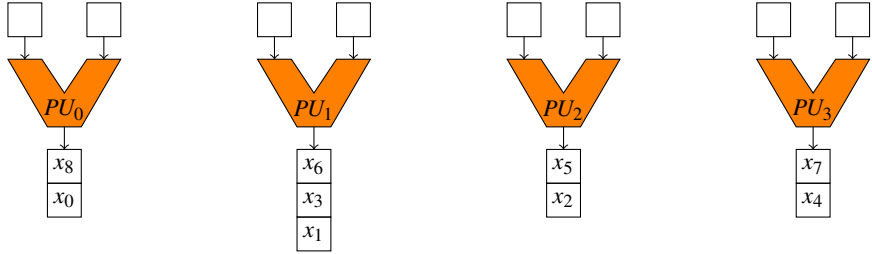

The answer can be easily computed by our ASP encoding if we switch the priorities of the two minimize statements for PUs and execution time. One out of the 6912 possible minimal schedules (288 with symmetry breaking) for our example DAG from \autoref{fig:example_bb} can be found in \autoref{fig:example_bb_solutions_exec_minimization}. As it can be seen, our initial basic block can be scheduled on 4 PUs with the minimal cost of 3 (as this was the maximal depth in the initial DAG).

%% file: Chap5_OptimalFurthers.tex
\section{Enhanced Optimal Schedulings}
\label{chap:optMore}

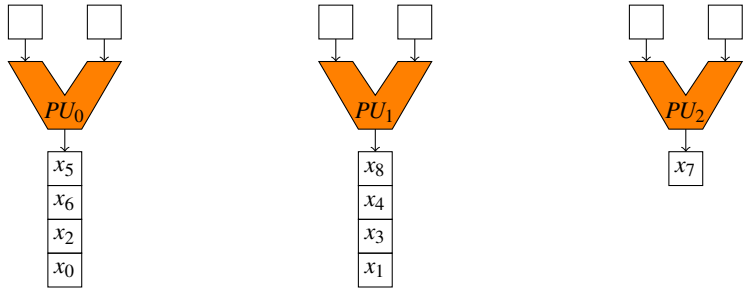
\begin{figure}[h]
\begin{subfigure}[t]{0.3\textwidth}
%Solution PU0
\begin{tikzpicture}[scale=0.3,baseline=32]
\draw  (0.5,3) rectangle (-1,4.5);
\node at (-0.25,3.75) {};
\draw  (-3,3) rectangle (-4.5,4.5);
\node at (-3.75,3.75) {};

\draw [fill=orange] (-2.75,-1) -- (-1.25,-1) -- (0.5,2) -- (-1,2) -- (-2,0.5) -- (-3,2) -- (-4.5,2) -- cycle;
\node at (-2,-0.25) {$PU_0$};

\draw  (-2.75,-2) rectangle (-1.25,-3.5);
\node at (-2,-2.75) {$x_5$};

\draw  (-2.75,-3.5) rectangle (-1.25,-5);
\node at (-2,-4.25) {$x_6$};

\draw  (-2.75,-5) rectangle (-1.25,-6.5);
\node at (-2,-5.75) {$x_2$};

\draw  (-2.75,-6.5) rectangle (-1.25,-8);
\node at (-2,-7.25) {$x_0$};

\draw [->] (-3.75,3) -- (-3.75,2);
\draw [->] (-0.25,3) -- (-0.25,2);
\draw [->] (-2,-1) -- (-2,-2);
\end{tikzpicture}
\end{subfigure}
%Solution PU1
\begin{subfigure}[t]{0.3\textwidth}
\begin{tikzpicture}[scale=0.3,baseline=32]
\draw  (0.5,3) rectangle (-1,4.5);
\node at (-0.25,3.75) {};
\draw  (-3,3) rectangle (-4.5,4.5);
\node at (-3.75,3.75) {};

\draw [fill=orange] (-2.75,-1) -- (-1.25,-1) -- (0.5,2) -- (-1,2) -- (-2,0.5) -- (-3,2) -- (-4.5,2) -- cycle;
\node at (-2,-0.25) {$PU_1$};

\draw  (-2.75,-2) rectangle (-1.25,-3.5);
\node at (-2,-2.75) {$x_8$};

\draw  (-2.75,-3.5) rectangle (-1.25,-5);
\node at (-2,-4.25) {$x_4$};

\draw  (-2.75,-5) rectangle (-1.25,-6.5);
\node at (-2,-5.75) {$x_3$};

\draw  (-2.75,-6.5) rectangle (-1.25,-8);
\node at (-2,-7.25) {$x_1$};

\draw [->] (-3.75,3) -- (-3.75,2);
\draw [->] (-0.25,3) -- (-0.25,2);
\draw [->] (-2,-1) -- (-2,-2);
\end{tikzpicture}
\end{subfigure}
%Solution PU2
\begin{subfigure}[t]{0.3\textwidth}
\begin{tikzpicture}[scale=0.3,baseline=32]
\draw  (0.5,3) rectangle (-1,4.5);
\node at (-0.25,3.75) {};
\draw  (-3,3) rectangle (-4.5,4.5);
\node at (-3.75,3.75) {};

\draw [fill=orange] (-2.75,-1) -- (-1.25,-1) -- (0.5,2) -- (-1,2) -- (-2,0.5) -- (-3,2) -- (-4.5,2) -- cycle;
\node at (-2,-0.25) {$PU_2$};

\draw  (-2.75,-2) rectangle (-1.25,-3.5);
\node at (-2,-2.75) {$x_7$};

\draw [->] (-3.75,3) -- (-3.75,2);
\draw [->] (-0.25,3) -- (-0.25,2);
\draw [->] (-2,-1) -- (-2,-2);
\end{tikzpicture}
\end{subfigure}
\vspace{1ex}
\caption{One example out of 8088 (without symmetries: 1348) to minimize the amount of PUs with an upper bound to 4 execution steps from the example basic block shown in \autoref{fig:example_bb}}
\label{fig:example_bb_solutions_boundary}
\end{figure}

So far, we have been able to answer the questions about the minimal execution time, and the minimal execution time for the least possible amount of PUs. More realistic is to consider both resource and time constraints together instead of after each other. For example, we may wish to determine the minimal number of PUs to execute a given program/basic block within $N$ steps. As another example, we may wish to determine the minimal execution time of a given program with a given number of PUs. To solve these problems, only minor changes are required in our ASP encoding which shows the benefits of using ASP. As the $maximalCost$ is already available in our model, we can answer the first question by adding just one constraint that states that the cost does not exceed a given constant for maximal executions.
\programmath
\[
\begin{array}{rcl}
    &:-&maximalCost(N),~N>max\_execution.
\end{array}
\]
\unprogrammath
If we execute this ASP encoding on our example basic block from \autoref{fig:example_bb} together with an upper bound of execution time 4, results as shown in \autoref{fig:example_bb_solutions_boundary} can be found. We can see that we need at least 3 PUs if it is needed to have a real-time behaviour of at most 4 steps.

Second, we can introduce a resource limitation of PUs by restricting the answers to not exceed our PU limitation:
\programmath
\[
\begin{array}{rcl}
    &:-&amountPUs(N),~N>pus\_available.
\end{array}
\]
\unprogrammath

%% file: Chap6_Benchmarks.tex
\section{Comparison with SAT and SMT/Benchmarks}
\label{chap:benchmarks}
We used the same random basic block generator as in \cite{BhSc16} to generate basic blocks whose DAG representations have a given size $n$ and a given number of levels $l$ \footnote{The tools are available at \href{https://es.cs.uni-kl.de/tools/teaching/}{https://es.cs.uni-kl.de/tools/teaching/}}. For every  pair (node,level), $1000$ basic blocks were generated. Schedules are derived for each basic block using the ASP solver Clingo with the following optimization criteria: (1) Minimize number of PUs required to execute the basic block without overhead. This is used for a comparison with the SAT encoding in \cite{BhSc16}. (2) Minimize the number of PUs and then minimize the time needed to execute the basic block on a SCAD machine without overhead. This is used for a comparison with the SMT encoding in \cite{BhSc17}. All the runs are performed on an Intel Core-i5 (4 x 2.67 GHz) desktop computer with 8 GB RAM running Ubuntu 14.04, the same machine that has been used for the SMT and the SAT encoding. We used the newest Clingo version 5.2.2 for the experiments, our own SAT solver, and Microsoft's Z3 SMT solver.

\begin{figure}
\begin{subfigure}[b]{0.475\textwidth}
\begin{center}
\includegraphics[width=\linewidth]{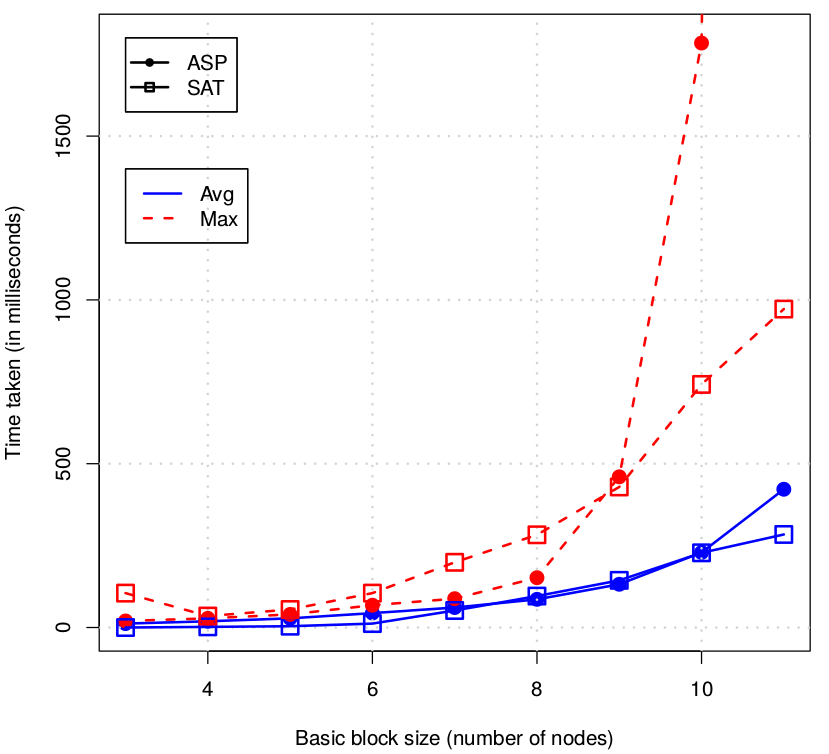}
\end{center}
\end{subfigure}
\quad
\begin{subfigure}[b]{0.475\textwidth}
\includegraphics[width=\linewidth]{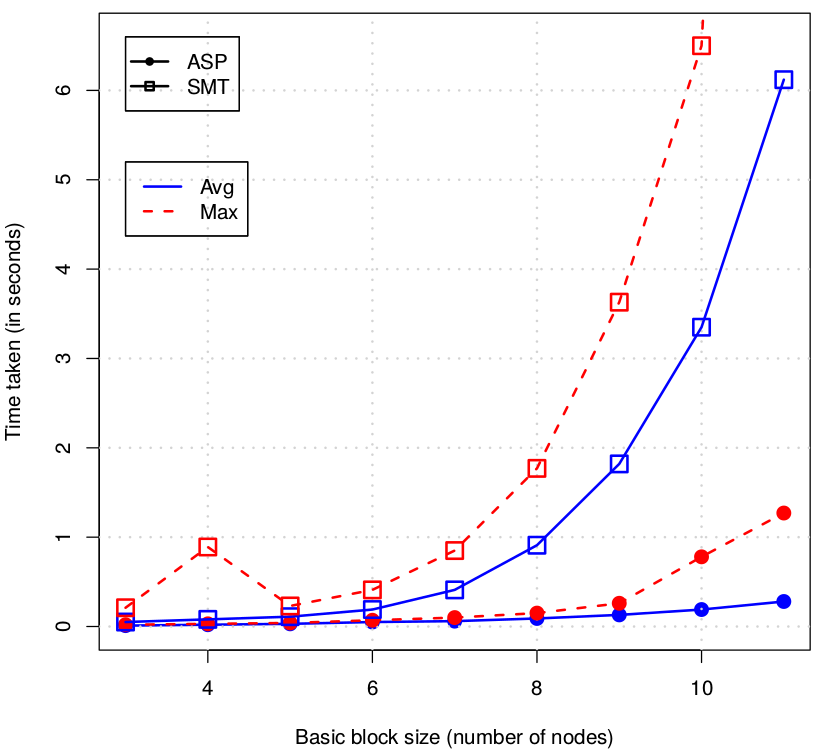}
\end{subfigure}
\vspace{1ex}
\caption{(a) Average and maximum time required by SAT and ASP solvers to derive resource constrained schedules for programs of different sizes. (b) Average and maximum time taken by SMT and ASP solvers to derive time constrained schedules for programs of different sizes.}
\label{fig:satSmtExps}
\end{figure}

The resulting execution times of Clingo are given in \autoref{fig:satSmtExps}, in comparison to SAT (a) and SMT (b). For the comparison with the SAT encoding, only PU optimization has been included in the ASP template to perform a fair comparison. We could not detect any significant differences between the average execution times. This is a quite good result because Clingo produced \emph{all} possible answer sets, whereas the SAT encoding only produced one satisfying model. However, there seems to be the trend that ASP will lead to higher maximal CPU times compared to SAT.

Because the SMT encoding was searching the minimal execution time on minimal numbers of PUs, minimization of the execution time has been included to the ASP encoding with second preference. It was quite interesting to see that the CPU times needed by Clingo did even decrease compared to the version without execution time minimization. As minimizing the execution time of a schedule was straightforwardly added to ASP, this is directly visible in the result, as the ASP encoding outperforms the SMT encoding by far. We could easily compute schedules with minimal execution times up to 20 nodes without having memory, time or other resource problems.

%% file: Chap7_RelatedWork.tex
%-------------------------------------------------------------------------------
\section{Related Work} \label{chap:relWork}
%-------------------------------------------------------------------------------

Answer set programming is known to be well-suited for solving scheduling problems (see e.g., \cite{DoMa17} and \cite{Bald11}) since optimization statements have found their way into ASP \cite{BrNT03}. Optimal scheduling for multicore processors has been investigated in \cite{KuDe09} using path cost minimization algorithms for weighted graphs. It allows mapping tasks to CPUs, if they are split before into independent tasks. The algorithms presented are directly usable for quadcore and dualcore CPUs and need manual changes for others. 

Symbolic system synthesis has been performed in \cite{IMBG09} and \cite{AGSH13a}. These synthesized systems are build to match best with a given problem instance, and find the most suitable system design. In our paper, we already have a system design, and want to find the best scheduling for this design.

There is a very interesting approach generating the best operations which lead to execution time minimization in \cite{BCVF06} and \cite{CBVF09}. Although the workflow proposed there is focusing on optimal code synthesis, and not finding an optimal schedule, it is still related to this paper, especially because they introduced a levelized architecture for reducing the search space. The basic idea is kept in mind for future optimizations of our work.
The work shown in \cite{AnPP12} focused in a similar way on the generation of concurrent programs described by structural and behavioral properties.

%% file: Chap8_Conclusion.tex
\section{Conclusions}
\label{chap:conclusion}

This paper introduced and evaluated the optimal code generation of basic blocks for SCAD machines using ASP. The flexibility of ASP can be used to adapt existing encodings to solve  new resource and time constrained optimization problems. The use of ASP solvers is competitive to SAT solvers, and outperforms the previously used SMT solvers. Additionally, ASP not only computes one optimal schedule, but all optimal solutions so that it is now possible to study their properties.